\documentclass[prb,twocolumn,showpacs,preprintnumbers]{revtex4}
\usepackage{amssymb}
\usepackage{amsmath}
\usepackage{epsfig}
\usepackage{graphicx}
\usepackage{dcolumn}
\usepackage{bm}

\begin{document}

\title{Bose-Einstein condensation in multilayers}
\author{P. Salas$^{1,2}$, M. Fortes$^2$, M. de Llano$^3$, F.J. Sevilla$^2$
and M.A. Sol\'{\i}s$^2$}

\address{$^1$Posgrado en Ciencias e Ingenier\'{\i}a de Materiales, UNAM, M\'exico, D.F., MEXICO\\
$^2$Instituto de F\'{\i}sica, UNAM, M\'exico, D.F., MEXICO \\
$^3$Instituto de Investigaciones en Materiales, UNAM, M\'exico, D.F., MEXICO}

\begin{abstract}
The critical BEC temperature $T_{c}$ of a non interacting boson gas in a
layered structure like those of cuprate superconductors is shown to have a
minimum $T_{c,m}$, at a characteristic separation between planes $a_{m}$. It
is shown that for $a<a_{m}$, $T_{c}$ increases monotonically back up to the
ideal Bose gas $T_{0}$ suggesting that a reduction in the separation between
planes, as happens when one increases the pressure in a cuprate, leads to an
increase in the critical temperature. 
For finite plane separation and penetrability the specific heat as a
function of temperature shows two novel crests connected by a ridge 
in addition to the well-known BEC peak at $T_{c}$ associated with the 3D
behavior of the gas. For completely impenetrable planes the model reduces to
many disconnected infinite slabs for which just one hump survives becoming a
peak only when the slab widths are infinite.
\end{abstract}

\maketitle

Since London first suggested \cite{London} that superfluidity in liquid $^{4}$He
might well be a manifestation of Bose-Einstein condensation (BEC) of the
helium atoms before interatomic interactions are ``switched
on," BEC\ in layered systems began to be studied to understand
helium films \cite{Brewer,Gasparini,Kinosita}. The discovery of high-$T_{c}$
superconductivity stimulated renewed interest in compounds with layered
structures \cite{Logvenov} in which a BEC mechanism seems to be an essential
feature to explain high critical temperatures \cite{WK}. The so-called
\textquotedblleft Uemura plot\textquotedblright\ \cite{uemura04} of data
from muon-spin relaxation ($\mu $SR), neutron and Raman scattering, and
angle-resolved photoemission (ARPES) measurements exhibits $T_{c}$ \textit{vs%
} Fermi temperatures $T_{F}\equiv E_{F}/k_{B}$ where $E_{F}$ the Fermi
energy and $k_{B}$ the Boltzmann constant. Empirical $T_{c}$s\ of many
cuprates straddle a line \textit{parallel} to the Uemura-plot diagonal line
associated with the simple BEC formula $T_{0}\simeq 3.31\hbar
^{2}n_{B}^{2/3}/mk_{B}\simeq 0.218T_{F}$ corresponding to an ideal gas of
bosons of mass $m=2m^{\ast }$ and number density $n_{B}=n_{s}/2$ where $%
m^{\ast }$ is the individual-charge-carrier effective mass and $n_{s}$ their
number density. The parallel line of data is \textit{shifted down} from $%
T_{0}$\ by a factor of $4$-$5$. This has been judged \cite{Uemura06}\ a
\textquotedblleft fundamental importance of the BEC concept in
cuprates.\textquotedblright\ In addition, the possibility of creating BECs
or superfluidity of ultracold fermions \cite{SF} in optical lattices \cite%
{OpLattice}, along with the expected observation of BEC of excitons
(electron-hole pairs) in semiconductors \cite{excitons}, have further
revived theoretical and experimental \cite{Clade, Roati} efforts to better
understand the behavior of quantum gases in layered geometries.

Most models based on layered structures \cite{WK,BF51} simulating quasi-2D
high-$T_{c}$ superconductors, or other to study BEC \cite{HR,Trifea,Hong},
rely on a single-boson hopping interaction term producing nearest-interlayer
couplings in one spatial dimension while moving freely in the other two
directions. The energy spectrum is typically of the form $\epsilon _{\mathbf{%
k}}=\hbar ^{2}(k_{x}^{2}+k_{y}^{2})/2m+\epsilon _{k_{z}}$ with $\epsilon
_{k_{z}}=(\hbar ^{2}/{Ma^{2}})(1-\cos k_{z}a)$ where $a$ is the plane
separation and the constant $\hbar ^{2}/{Ma^{2}}$ is a measure of the
bosonic Cooper pair hopping probability between planes. In the case of 
CuO$_{2}$ planes in cuprate superconductors a boson effective mass $M$ in the 
$z$-direction is used with $m$ the mass in the $x$-, $y$-directions while $%
M>m$ reflects cuprate anisotropy. For $k_{z}a\ll 1$ one recovers the
expected result $\epsilon _{k_{z}}=\hbar ^{2}k_{z}^{2}/2{M}$.

Motivated by possible applications to quasi-2D superconductors or to $^{4}$He 
films, in this Brief Report we discuss critical temperature $T_{c}$\ and
the specific heat $C_{V}$ results in a non-interacting boson gas of $N$
particles within an infinite multilayered structure of equally-spaced planes
of variable penetrability stacked in the $z$-direction. 

The system of parallel planes is simulated with an external periodic delta
potential of the 1D Kronig-Penney (KP) type along the $z$-direction while
bosons are free in the two remaining directions. Two important differences
with previous studies \cite{WK,HR,Trifea,Hong} are: a) the bosons are
allowed to move over all space through permeable planes instead of
constraining them to move only over the plane surfaces, and b) the
anisotropy expected from a layered material is introduced naturally by the
delta KP potential. In addition, the delta strengths avoid the artificial
introduction of large masses and hopping parameters in the $z$-direction to
tune the mobility or tunneling across planes. At very low energies the KP
potential reduces to the energy expression $\epsilon _{k_{z}}$ used in Refs.%
\cite{WK,HR,Trifea,Hong} which only takes into account the lowest allowed
energy band. If $T_{0}$ is the critical BEC temperature of the free 3D boson
gas and the associated thermal wavelength is $\lambda _{0}\equiv h/\sqrt{%
2\pi mk_{B}T_{0}}$ a dimensionless \textquotedblleft
impenetrability\textquotedblright\ $P_{0}\equiv P\lambda _{0}\geqslant 0$ of
layers is introduced in terms of the KP penetrability parameter $P$ [see (%
\ref{KPdelta}) below]$.$ Thus, $P_{0}=0$ implies perfect layer transparency
and $P_{0}=\infty $ fully opaque layers. In the latter limit, our model
reduces to that of an infinite number of uncoupled slabs of thickness $a$
and infinite lateral extent. The specific heat has been obtained in these
systems to model $^{4}$He thin-film properties \cite%
{Ziman,Goble,Pathria72,Greenspoon} and finite-size effects on BEC \cite%
{Pajkowski}. In these infinite slabs the BEC critical temperature is zero
but the specific heat shows a smooth maximum at a temperature that depends
on the slab thickness. Some authors \cite{Pathria72,Greenspoon,Pajkowski}
have associated this \textquotedblleft hump\textquotedblright\ with a BEC
signature. However, we show here that for finite $P_{0}$ this hump coexists
with a nonzero critical BEC temperature in addition to a second maximum and
when $P_{0}\rightarrow \infty $ the BEC temperature goes to zero whereas the
hump persists to become the familiar BEC with a \textit{cusped} sharp peak
only in the limit $a/\lambda _{0}\rightarrow \infty $.

The Schr\"{o}dinger equation for a boson of mass $m$ is separable in the $x$%
, $y$ and $z$-directions so that the single-particle energy as a function of
the wavevector $\mathbf{k=(}k_{x},k_{y},k_{z})$ is simply $\varepsilon _{%
\mathbf{k}}=\varepsilon _{k_{x}}+\varepsilon _{k_{y}}+\varepsilon _{k_{z}}$,
where $\varepsilon _{k_{x},k_{y}}=\hbar ^{2}k_{x,y}^{2}/2m$ with $%
k_{x,y}=2\pi n_{x,y}/L$ and $n_{x,y}=0,\pm 1,\pm 2,\cdots $, i.e., in the $x$
and $y$-directions particles are free and we assume periodic boundary
conditions in a box of size{\LARGE \ }$L$. In the $z$-direction the
particles are subject to the KP periodic delta (or \textquotedblleft Dirac
comb\textquotedblright ) potential in 1D \cite{KP} and the energies $%
\varepsilon _{k_{z}}$ are implicit in the transcendental equation 
\begin{equation}
Pa\frac{\sin \,\alpha a}{\alpha a}+\cos \,\alpha a=\cos \,k_{z}a
\label{KPdelta}
\end{equation}%
where $\alpha ^{2}\equiv 2m\varepsilon _{k_{z}}/\hbar ^{2}$ and $\hbar
^{2}P/m$ is the delta-interaction strength or equivalently, the layer
impenetrability parameter. The energy $\varepsilon _{k_{z},j}$ depends on
the parameters $P$ and $a$ where the band structure is labeled with the
index $j=1,2,...$ The trivial free-particle energy dispersion relation in
the $z$-direction, $\varepsilon _{k_{z},j}\rightarrow \hbar ^{2}k_{z}^{2}/2m,
$ is recovered in the limit $P\rightarrow 0$, while $P\rightarrow \infty $
yields $\sin (\alpha a)\rightarrow 0$ which corresponds to a system of
confined bosons inside a semi-infinite slab of width $a,$ a situation
extensively discussed in the literature (see \cite{Greenspoon} and
references therein). For $\alpha a\ll 1$, (\ref{KPdelta}) can be expanded as%
\begin{equation*}
\varepsilon _{k_{z}}\simeq \frac{\hbar ^{2}}{2ma^{2}}\frac{Pa+(1-\cos k_{z}a)%
}{1/2+Pa/6}\underset{Pa\ll 1}{\longrightarrow }\frac{\hbar ^{2}}{ma^{2}}%
(1-\cos k_{z}a)
\end{equation*}%
where the last expression is the dispersion relation used in \cite%
{WK,HR,Hong}. For finite $P$, bosons tunnel through the layers as a more
realistic model would demand.

The grand potential $\Omega (T,V,\mu )$\ $=-pV$ (with $p$ the pressure, $V$
the volume and $\mu $ the chemical potential) leads to all the system
thermodynamic properties. In our multilayered system it becomes, after
integrating over $k_{x}$ and $k_{y}$, 
\begin{equation}
\Omega (T,V,\mu )=\Omega _{0}-\frac{1}{\beta ^{2}}\frac{Vm}{\left( 2\pi
\right) ^{2}\hbar ^{2}}\sum_{j=1}^{\infty }{\int_{-\pi /a}^{\pi /a}dk_{z}}%
g_{2}(ze^{-\beta \varepsilon _{k_{z}j}})  \label{TGP}
\end{equation}%
where $z\equiv e^{\beta \mu }$ is the fugacity, $\beta \equiv 1/k_{B}T$, and
we have explicitly separated the contribution $\Omega _{0}\equiv k_{B}T\ln
[1-ze^{-\beta \varepsilon _{0}}]$ of the lowest-energy state\ $\varepsilon
_{0}$. Here $g_{\sigma }(t)$ stands for the Bose function \cite{Pathrialibro}%
. The integration is over the first Brillouin zone and the sum is over the
allowed bands. Note that the ground-state energy $\varepsilon _{0}$ depends
on both $P$ and $a$ in (\ref{KPdelta})$.$

The average number of particles $N=-\left( \partial \Omega /\partial \mu
\right) _{T,V}$ for a free boson gas in the multilayered configuration is
then%
\begin{equation}
N=N_{0}(T)-\frac{Vm}{\left( 2\pi \right) ^{2}\hbar ^{2}}\frac{1}{\beta }{%
\sum_{j=1}^{\infty }\int_{-\pi /a}^{\pi /a}dk_{z}}\ln [1-ze^{-\beta
\varepsilon _{k_{z}j}}]  \label{num2}
\end{equation}%
where $N_{0}(T)\equiv \lbrack z^{-1}e^{\beta \varepsilon _{0}}-1]^{-1}$ is
the number of particles condensed in the ground state $\varepsilon _{0}$
while the last term in (\ref{num2}) is the number of particles in excited
states. As usual, the BEC critical temperature $T_{c}$ is obtained by
finding the temperature below which the fractional number of particles in
the ground state just ceases to be negligible upon cooling and the chemical
potential reaches its maximum value $\varepsilon _{0}>0$ making the fugacity
equal $z_{0}=e^{\beta _{c}\varepsilon _{0}}$. For convenience and clarity we
use the same particle density $N/V$ in the layered system as that for the
free ideal Bose gas (IBG) with the BEC critical temperature $T_{0}.$ Thus $%
k_{B}T_{0}$ serves as an energy scale while $\lambda _{0}\equiv h/\sqrt{2\pi
mk_{B}T_{0}}$ provides a length scale.

\begin{figure}[tbh]
\centerline{\epsfig{file=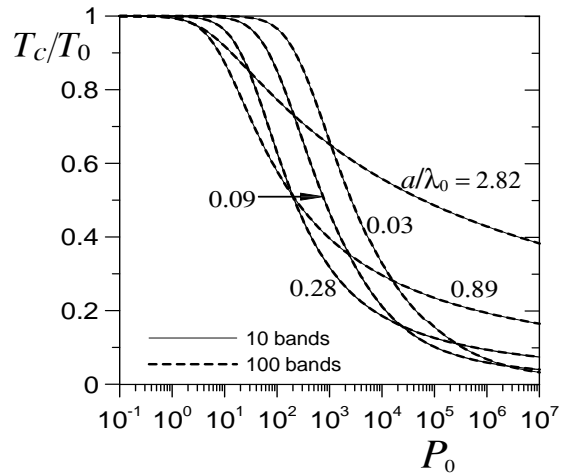,height=2.50in,width=3.0in}} 
\caption{Critical temperature $T_{c}$ in units of $T_{0}$ as a function of $%
P_{0}$ for different values of $a/\protect\lambda _{0}$.}
\label{fig:TcvsP0}
\end{figure}

The effects of the KP layers on BEC are shown in Fig. \ref{fig:TcvsP0} where 
$T_{c}/T_{0}$ decreases as $P_{0}$ increases for different values of $a.$
Note that the critical temperature is very sensitive to the number of energy
bands intervening in the numerical calculations. Here, it is sufficient to
include 10 bands to achieve convergence for the values of $a$ and $P_{0}$
shown in the figure. When $P_{0}=0$ (perfect barrier transparency) we
recover the IBG results. In the opposite limit $P_{0}\rightarrow \infty $
(impenetrable barriers) $T_{c}/T_{0}$ vanishes although strictly speaking
the model does not become a true two-dimensional system.

\begin{figure}[tbh]
\centerline{\epsfig{file=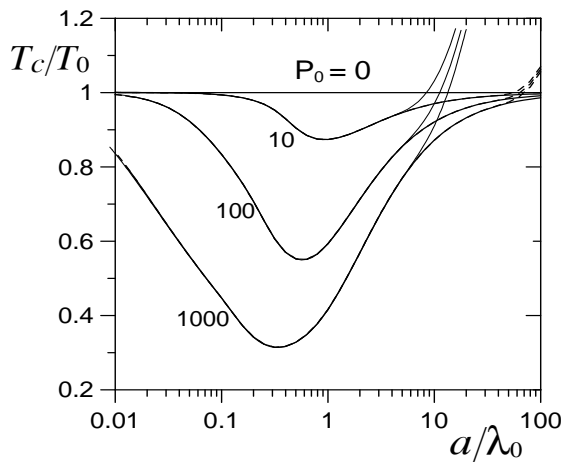,height=2.5in,width=3.0in}}. 
\vspace{-0.5cm}
\caption{Critical temperature as a function of $a/\protect\lambda _{0}$ for
different values of $P_{0}$. For $a\gg \protect\lambda _{0}$ the critical
temperature is very sensitive to the number of energy bands used in the
numerical calculations. Thin, dashed and normal lines correspond to 10, 100
and 1000 bands, respectively. To attain convergence we need to consider more
than ten bands (thin line) which contrasts with the behavior in the $a<%
\protect\lambda _{0}$ where the inclusion of few bands is enough to obtain
convergence}
\label{fig:TcvsG}
\end{figure}

A novel result shown in Fig. \ref{fig:TcvsG} is that for finite $P_{0},$ $%
T_{c}/T_{0}$ diminishes from $1$ down to a minimum value as the plane
separation $a$\ decreases. For $P_{0}$ spanning at least five orders of
magnitude (see Table \ref{tab:table1}) the critical-temperature minima show
a linear dependence with 
\ $a_{m}$ so that $T_{c,m}=(T_{0}/\lambda _{0})a_{m}$ or ($\lambda
_{0}/\lambda _{c,m})^{2}=a_{m}/\lambda _{0}$, where the subscript $m$ means minimum and $\lambda _{c,m}$ is the thermal wave length associated to $T_{c,m}$. Since 
$T_{c,m}\rightarrow 0$ and $a_{m}\rightarrow 0$ for $P_{0}\rightarrow \infty 
$ this linear relation seems to hold for all $P_{0}\gtrsim 10$. For $P_{0}$%
$\lesssim 10$, $T_{c,m}/T_{0}$ approaches $1$ while 
$a_{m}\rightarrow \infty $. Surprisingly, further reduction in 
$a$ brings an \textit{increase} in $T_{c}/T_{0}$ which asymptotically
reaches unity. This can be understood from the KP dispersion relation (\ref%
{KPdelta}) since for small values of $a$ sin$(\alpha a)/\alpha a\simeq 1$
and $\varepsilon _{k_{z},j}\simeq \hbar ^{2}k_{z}^{2}/2m$ so
that one recovers the 3D IBG regime when $a\rightarrow 0$, as expected. We
also note in Fig. \ref{fig:TcvsG} that $T_{c}/T_{0}$ is not symmetric wrt
the minimum value. For $a/\lambda _{0}\ll 1$, $T_{c}/T_{0}$ increases slower
than for $a/\lambda _{0}\gg 1$ which reflects the plane influence on the
bosons. We also note that for $a/\lambda _{0}\gg 1$ one must include more
than ten energy bands to achieve convergence in the numerical calculations.

\begin{table}[tbp]
\caption{Values of plane separation $a_{m}$ for which the critical
temperature reaches its minimum for different $P_{0}.$}
\label{tab:table1}%
\begin{ruledtabular}
\begin{tabular}{cccccc}
$P_{0}$ & 10 & 10$^2$ & 10$^3$ &10$^4$ & 10$^5$\\
\hline
$a_{m}/\lambda_{0}$  &   0.874 & 0.550 & 0.314 & 0.177 & 0.100 \\
$T_{cm}/T_{0}$  & 0.927 & 0.572 & 0.337&  0.191& 0.107   \\
\end{tabular}
\end{ruledtabular}
\end{table}
%
%
%
%
%
%
%

The specific heat $C_{V}=-T\left[ {\partial }^{2}\Omega /\partial T^{2}%
\right] _{V,\mu }$ is 
\begin{eqnarray}
\frac{C_{V}}{Nk_{B}} &=&\frac{Vm}{N(2\pi )^{2}\hbar ^{2}}\sum_{j=1}^{\infty }%
{\int_{-\pi /a}^{\pi /a}dk_{z}}\left[ 2k_{B}T\,g_{2}(ze^{-\beta \varepsilon
_{k_{z}j}})\right.   \notag \\
&&\hspace{-1.5cm}-\ln (1-ze^{-\beta \varepsilon _{k_{z}j}})\left[
2\,\varepsilon _{k_{z}j}-\mu -\varepsilon _{0}+T\frac{\partial \mu }{%
\partial T}\right]   \notag \\
&&\hspace{-1.5cm}\left. +\frac{1}{k_{B}T}\,\frac{(\varepsilon
_{k_{z}j}-\varepsilon _{0})[\varepsilon _{k_{z}j}-\mu +T\frac{\partial \mu }{%
\partial T}]}{z^{-1}e^{\beta \varepsilon _{k_{z}j}}-1}\right]   \label{Cv2}
\end{eqnarray}%
%
%
%
%
where 
\begin{equation}
T(d\mu /dT)=\frac{-\dfrac{N}{V}\dfrac{(2\pi )^{2}\hbar ^{2}}{m}%
-\sum\limits_{j=1}^{\infty }\int\limits_{-\pi /a}^{\pi /a}dk_{z}\dfrac{%
\varepsilon _{k_{z}j}-\mu }{z^{-1}e^{\beta \varepsilon _{k_{z}j}}-1}}{%
\sum\limits_{j=1}^{\infty }\int\limits_{-\pi /a}^{\pi /a}dk_{z}\left[
z^{-1}e^{\beta \varepsilon _{k_{z}j}}-1\right] ^{-1}}  \notag
\end{equation}%
%
%
%
%
\begin{figure}[tbh]
\centerline{\epsfig{file=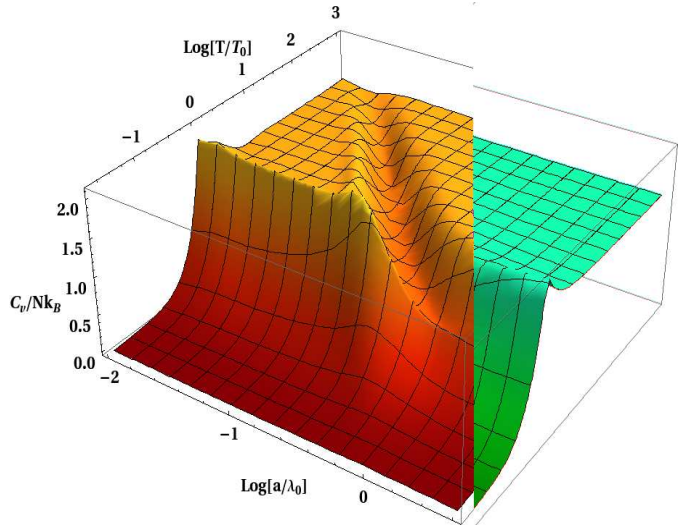,height=3.5in,width=3 in,angle=90}} 
\caption{3D plot of the specific heat as function of temperature and plane
separation $a$ for $P_{0}=100$.}
\label{fig:CvOmar1}
\end{figure}
\vspace{-0.7cm} 
\begin{figure}[tbh]
\centerline{%
\epsfig{file=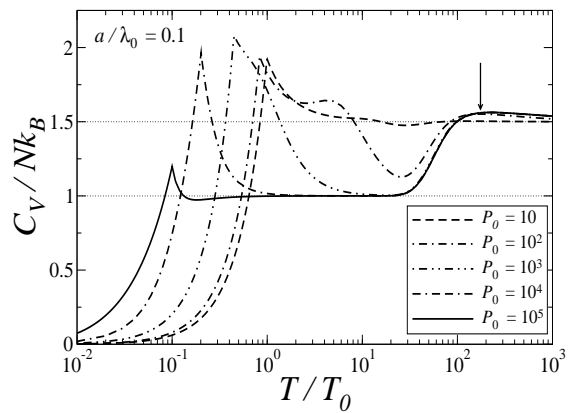,height=3.0in,width=2.5in,angle=-90}} 
\caption{Specific heat per particle as a function of temperature for several 
$P_{0}$ and $a/\protect\lambda _{0}=0.1$. The two-dimensional character of
the layered structure is easily seen as $P_{0}$ increases. The arrow
indicates the initiation of the IBG regime}
\label{fig:CvFJ}
\end{figure}
%
%
%
For $T<T_{c}$ the chemical potential $\mu =\varepsilon _{0}$ is a constant
so that $\partial \mu /\partial T=0$ which simplifies the equation for $%
C_{V}/Nk_{B}$. %

Numerical results are summarized in Fig. \ref{fig:CvOmar1} of $C_{V}/Nk_{B}$
vs. $\log T/T_{0}$ and vs. $\log a/\lambda _{0}$ for an intermediate
dimensionless penetrability $P_{0}=100.$ In contrast with the IBG case, in
addition to the familiar BEC peak for finite $P_{0}$ two humps appear that
are connected by a ridge when $T>T_{c}$; ridge heights depend on the plane
separation. 
When $a>a_{m}$ only one hump survives and becomes the BEC cusped peak in the
limit $a\rightarrow \infty $ independently of $P_{0}$. Our model shows the
distinct nature of the ever present BEC critical temperature plus a \textit{%
second} characteristic temperature where $C_{V}/Nk_{B}$ is maximum. In
previous models (here equivalent to $P_{0}\rightarrow \infty $) \cite%
{Greenspoon,Pajkowski} this latter temperature is identified as a signature
of the BEC critical temperature. However, our results indicate that this
characteristic temperature should be associated with a different feature,
namely an accumulation of bosons \cite{Osborne} in the lowest energy levels.

While $C_{V}/Nk_{B}$ of the IBG decreases monotonically for $T>T_{c}\equiv
T_{0}$, in the present model there is a region defined by $a\lesssim a_{m}$
where $C_{V}/Nk_{B}$ varies non-monotonically with $T$ %
having a set of local minima (the ridge in Fig. \ref{fig:CvOmar1}) where $%
\lambda \simeq 2a$. In this case, the single-particle (thermal) average
wavefunction nodes coincide with the plane locations resulting in the
particle motion freezing out in the $z$-direction. 
In particular, for $a=a_{m}$ the minimum of $C_{V}(T)/Nk_{B}$ attains its
lowest value $\simeq 1$ so that the \textquotedblleft
ridge\textquotedblright\ reveals a closer resemblance to a 2D system. As $%
P_{0}$ increases this 2D-like feature dominates as the ridge levels out over
a broader temperature region (see Fig. \ref{fig:CvFJ} for $a/\lambda _{0}=0.1
$). In addition, for temperatures below $T_{c}$ the value $\lambda \simeq 2a$
marks a functional crossoever of specific heat from the standard 3D IBG $%
C_{V}\propto T^{3/2}$ to the 2D linear $C_{V}\propto T$ behavior. 
For sufficiently high temperatures, layer effects are negligible regardless
of separation $a$ or of penetrability $P_{0}$. In this regime $%
C_{V}(T)/Nk_{B}$ approaches the IBG value. Since $C_{V}$ attains this value
at the same $T$ (implying the same $\lambda $) 
we can identify an $a$-dependent correlation length that manifests itself
when $\lambda \simeq 0.8a$ as marked by the arrow in Fig. \ref{fig:CvFJ} for 
$a/\lambda _{0}=0.1$ and $T/T_{0}\simeq 170K$.

%
%

To conclude, we have shown that the BEC-like critical temperature as well as
the specific heat of a boson gas between penetrable and periodically-spaced
planes change drastically compared to previous models based on hopping,
Hubbard \cite{WK} or slab \cite{Greenspoon} configurations, or when there is
no confining geometrical structure whatsoever. Even though analytical
approximations for the boson dispersion relation at low energies commonly
used to describe layered systems have proved to be a useful guide to give
more accurate results, they are significantly improved when the more
accurate dispersion relation including more than one energy band is used.
The appearance of two characteristic temperatures associated with the maxima
of $C_{V}(T)$ in addition to the BEC transition temperature due to the
confinement parameters may explain the variation of $T_{c}$ with pressure in
high-$T_{c}$ superconductors the transition is related with a BEC-like
phenomenon. In the limit $P_{0}\rightarrow \infty $ (opaque or
perfectly-decoupled infinite slabs of width $a$) the BEC-like transition
temperature vanishes leaving only one maximum which becomes the usual
cusped-peak BEC transition only when $a\gg \lambda _{0}$.

We acknowledge partial support from grants PAPIIT IN114708 and IN106908.
F.J.S. acknowledges partial support from Conacyt-SNI-I-89774. We thank O.A.
Rodr\'{\i}guez for assistance in preparing Fig. \ref{fig:CvOmar1}.

\end{document}